\documentstyle [12pt] {article}

\catcode`@=12
\begin{document}
\thispagestyle{empty}
\begin{flushright} UCRHEP-T259\\July 1999\
\end{flushright}
\vspace{0.5in}
\begin{center}
{\Large \bf Stability of Neutrino Mass Degeneracy\\}
\vspace{1.8in}
{\bf Ernest Ma\\}
\vspace{0.3in}
{\sl Department of Physics\\}
{\sl University of California\\}
{\sl Riverside, California 92521\\}
\vspace{1.8in}
\end{center}
\begin{abstract}\ 
Two neutrinos of Majorana masses $m_{1,2}$ with mixing angle $\theta$ 
are unstable against radiative corrections in the limit $m_1 = m_2$, but 
are stable for $m_1 = -m_2$ (i.e. opposite $CP$ eigenstates) with $\theta 
= 45^\circ$ which corresponds to an additional symmetry.
\end{abstract}
\newpage
\baselineskip 24pt
Pick two neutrinos, say $\nu_e$ and $\nu_\mu$.  Assume their mass eigenstates 
to be
\begin{equation}
\nu_1 = \nu_e \cos \theta - \nu_\mu \sin \theta, ~~~ \nu_2 = \nu_e \sin \theta 
+ \nu_\mu \cos \theta,
\end{equation}
with eigenvalues $m_1$ and $m_2$ respectively.  Neutrino oscillations may 
then occur\cite{1,2,3} if both $\Delta m^2 = m_2^2 - m_1^2$ and $\sin^2 2 
\theta$ are nonzero.  However, it is entirely possible that the hierarchy
\begin{equation}
\Delta m^2 << m_{1,2}^2
\end{equation}
actually exists, so that the smallness of $\Delta m^2$ for neutrino 
oscillations does not necessarily preclude a much larger common mass for 
the two neutrinos.  In fact, this idea is often extended to all three 
neutrinos\cite{4,5}.  On the other hand, since the charged-lepton masses 
are all different, radiative corrections\cite{6,7} to $m_1$ and $m_2$ 
will tend to change $\Delta m^2$ as well as $\theta$.  This is especially 
important for the vacuum oscillation solution\cite{8} to the observed solar 
neutrino deficit\cite{2} which requires $\Delta m^2 \sim 10^{-10}$ eV$^2$ 
and $\sin^2 2 \theta \sim 1$. 
In the following I show that whereas the limit $m_1 = m_2$ is unstable 
against radiative corrections, the limit $m_1 = -m_2$ and $\theta = 45^\circ$ 
is stable because it is protected by an additional symmetry.  [A negative 
mass here means that the corresponding Majorana neutrino is odd under $CP$ 
after a $\gamma_5$ rotation to remove the minus sign.]

Consider the $2 \times 2$ mass matrix spanning $\nu_e$ and $\nu_\mu$:
\begin{equation}
{\cal M} = \left( \begin{array} {c@{\quad}c} A & B \\ B & C \end{array} 
\right).
\end{equation}
It has eigenvalues
\begin{equation}
m_{1,2} = {1 \over 2} (C+A) \mp {1 \over 2} \sqrt {(C-A)^2 + 4B^2}
\end{equation}
where
\begin{eqnarray}
A &=& m_1 \cos^2 \theta + m_2 \sin^2 \theta, \\ 
B &=& (m_2 - m_1) \sin \theta \cos \theta, \\ 
C &=& m_1 \sin^2 \theta + m_2 \cos^2 \theta.
\end{eqnarray}
The mixing angle $\theta$ is related to $\cal M$ according to
\begin{equation}
\tan \theta = {2B \over (C-A) + \sqrt {(C-A)^2+4B^2}},
\end{equation}
and
\begin{equation}
\Delta m^2 = (C+A) \sqrt {(C-A)^2+4B^2}.
\end{equation}
In the above, I have used the convention $m_2 > |m_1|$ and $0 \leq \theta 
\leq 45^\circ$.

With radiative corrections, the mass matrix is changed:
\begin{equation}
A \to A (1 + 2 \delta_e), ~~~ B \to B (1 + \delta_e + \delta_\mu), ~~~ 
C \to C (1 + 2 \delta_\mu).
\end{equation}
If both $e$ and $\mu$ have only gauge interactions, then $\delta_e = 
\delta_\mu$ and $\cal M$ is simply renormalized by an overall factor, 
resulting in
\begin{equation}
\Delta m^2 \to \Delta m^2 (1 + 2 \delta)^2,
\end{equation}
and $\tan \theta$ is unchanged.  However, because $e$ and $\mu$ have Yukawa 
interactions proportional to their masses, nontrivial changes do occur in 
$\cal M$.  Let
\begin{equation}
\delta = (\delta_\mu + \delta_e)/2, ~~~ \Delta \delta = \delta_\mu - 
\delta_e,
\end{equation}
then
\begin{equation}
\Delta m^2 \to [(m_2 + m_1)(1 + 2 \delta) + (m_2 - m_1) \Delta \delta 
\cos 2 \theta]~D,
\end{equation}
and
\begin{equation}
\tan \theta \to {(m_2 - m_1) \sin 2 \theta (1 + 2 \delta) \over (m_2 - m_1) 
\cos 2 \theta (1 + 2 \delta) + (m_2 + m_1) \Delta \delta + D},
\end{equation}
where
\begin{equation}
D = \sqrt {(m_2-m_1)^2(1+2\delta)^2 + 2\Delta m^2 (1+2\delta) \Delta \delta 
\cos 2 \theta + (m_2+m_1)^2 (\Delta \delta)^2}.
\end{equation}

There are two ways for $\Delta m^2$ to approach zero:
\begin{equation}
(1) ~~ m_2 - m_1 << m_2 + m_1 = 2m,
\end{equation}
and
\begin{equation}
(2) ~~ m_2 + m_1 << m_2 - m_1 = 2m.
\end{equation}
In Case (1),
\begin{equation}
D \simeq 2m \sqrt {\left( {\Delta m^2 \over 4 m^2} \right)^2 + 2 \left( 
{\Delta m^2 \over 4 m^2} \right) \Delta \delta \cos 2 \theta + (\Delta 
\delta)^2}.
\end{equation}
Hence if $\Delta \delta >> \Delta m^2/4m^2$, then
\begin{equation}
\Delta m^2 \to 4 m^2 \Delta \delta, ~~~ \tan \theta \to 0,
\end{equation}
i.e. this situation is unstable.  Of course, if $\Delta m^2/4m^2 >> \Delta 
\delta$, there is no problem.  For example, if $\Delta m^2 \sim 10^{-3}$ 
eV$^2$ for atmospheric neutrino oscillations\cite{1} and $m \sim 1$ eV, 
then this is easily satisfied.  The model-independent contribution to 
$\Delta \delta$ from the renormalization of the neutrino wavefunctions is 
\begin{equation}
\Delta \delta = - {G_F (m_\mu^2 - m_e^2) \over 16 \pi^2 \sqrt 2} \ln 
{\Lambda^2 \over m_W^2},
\end{equation}
where $\Lambda$ is the scale at which the original mass matrix $\cal M$ is 
defined.  Other model-dependent contributions\cite{6} to the mass terms 
themselves may be of the same order.  If $m_\mu$ is replaced by $m_\tau$ 
in Eq.~(20), $\Delta \delta$ is of order 10$^{-5}$.  In that case, only the 
small-angle matter-enhanced solution\cite{9} to the solar neutrino deficit 
appears to be stable\cite{7} for $m \sim 1$ eV.

In Case (2),
\begin{equation}
D \simeq 2m (1+2\delta) \left[ 1 + \left( {\Delta m^2 \over 4 m^2} \right) 
{\Delta \delta \cos 2 \theta \over (1+2\delta)} \right],
\end{equation}
hence
\begin{equation}
\Delta m^2 \to \Delta m^2 (1+2\delta)^2 + 4 m^2 \Delta \delta \cos 2 \theta 
(1+2\delta),
\end{equation}
and
\begin{equation}
\tan \theta \to \tan \theta \left[ 1 - \left( {\Delta m^2 \over 4 m^2} \right)
\Delta \delta \right].
\end{equation}
This means that $\theta$ is stable and that $\Delta m^2$ is also stable if 
$\cos 2 \theta \simeq 0$, i.e. $\theta \simeq 45^\circ$.  More precisely, 
the condition
\begin{equation}
\Delta \delta \cos 2 \theta << {\Delta m^2 \over 4 m^2}
\end{equation}
is required.

Whereas the general form of $\cal M$ given by Eq.~(3) has no special symmetry 
for the entire theory, the limit $m_1 = -m_2$ and $\theta = 45^\circ$, i.e.
\begin{equation}
{\cal M} = \left( \begin{array} {c@{\quad}c} 0 & m \\ m & 0 \end{array} 
\right)
\end{equation}
is a special case which allows the entire theory to have the additional 
global symmetry $L_e - L_\mu$.  Hence small deviations are protected against 
radiative corrections, as shown by Eqs.(22) and (23).

The zero $\nu_e - \nu_e$ entry of Eq.~(25) also has the well-known virtue of 
predicting an effective zero $\nu_e$ mass in neutrinoless double beta decay. 
This means that $m$ may be a few eV even though the above experimental upper 
limit\cite{10} is one order of magnitude less.  Hence neutrinos could be 
candidates for hot dark matter\cite{11} in this scenario.

In conclusion, neutrino mass degeneracy is theoretically viable and 
phenomenologically desirable provided that $m_1 \simeq -m_2$ and $\theta 
\simeq 45^\circ$.

\vspace{0.3in}
\begin{center} {ACKNOWLEDGEMENT}
\end{center}

I thank V. Berezinsky and J. W. F. Valle for discussions. 
This work was supported in part by the U.~S.~Department of Energy under 
Grant No.~DE-FG03-94ER40837.

\newpage
\bibliographystyle {unsrt}

\begin{thebibliography} {99}
\bibitem{1} Y. Fukuda {\it et al.}, Phys. Lett. {\bf B433}, 9 (1998); 
{\bf B436}, 33 (1998); Phys. Rev. Lett. {\bf 81}, 1562 (1998); {\bf 82}, 2644 
(1999).
\bibitem{2} R. Davis, Prog. Part. Nucl. Phys. {\bf 32}, 13 (1994);  P. 
Anselmann {\it et al.}, Phys. Lett. {\bf B357}, 237 (1995); {\bf B361}, 235 
(1996); J. N. Abdurashitov {\it et al.}, Phys. Lett. {\bf B328}, 234 (1994); 
Y. Fukuda {\it et al.}, Phys. Rev. Lett. {\bf 77}, 1683 (1996); {\bf 81}, 
1158 (1998); {\bf 82}, 1810 (1999); {\bf 82}, 2430 (1999).
\bibitem{3} C. Athanassopoulos {\it et al.}, Phys. Rev. Lett. {\bf 75}, 2650 
(1995); {\bf 77}, 3082 (1996); {\bf 81}, 1774 (1998).
\bibitem{4} D. Caldwell and R. N. Mohapatra, Phys. Rev. {\bf D48}, 3259 
(1993);  A. S. Joshipura, Z. Phys. {\bf C64}, 31 (1994); Phys. Rev. {\bf D51}, 
1321 (1995); P. Bamert and C. P. Burgess, Phys. Lett. {\bf B329}, 289 (1994); 
D.-G. Lee and R. N. Mohapatra, Phys. Lett. {\bf B329}, 463 (1994); A. 
Ioannisian and J. W. F. Valle, Phys. Lett. {\bf B332}, 93 (1994); A. Ghosal, 
Phys. Lett. {\bf B398}, 315 (1997); A. K. Ray and S. Sarkar, Phys. Rev. 
{\bf D58}, 055010 (1998); C. D. Carone and M. Sher, Phys. Lett. {\bf B420}, 
83 (1998); H. Fritzsch and Z. Xing, Phys. Lett. {\bf B440}, 313 (1998); U. 
Sarkar, Phys. Rev. {\bf D59}, 037302 (1999); G. C. Branco, M. N. Rebelo, 
and J. I. Silva-Marcos, Phys. Rev. Lett. {\bf 82}, 683 (1999).
\bibitem{5} F. Vissani, hep-ph/9708483; H. Georgi and S. L. Glashow, 
hep-ph/9808293; R. N. Mohapatra and S. Nussinov, Phys. Rev. {\bf D60}, 013002 
(1999); Y. L. Wu, hep-ph/9810491, 9901245, 9901320; C. Wetterich, Phys. Lett. 
{\bf B451}, 397 (1999); R. Barbieri, L. J. Hall, G. L. Kane, and G. G. Ross, 
hep-ph/9901228.
\bibitem{6} E. Ma, Phys. Lett. {\bf B456}, 48 (1999); {\bf B456}, 201 (1999); 
hep-ph/9902392.
\bibitem{7} J. Ellis and S. Lola, hep-ph/9904279; J. A. Casas, J. R. 
Espinosa, A. Ibarra, and I. Navarro, hep-ph/9904395, 9905381, 9906281; 
R. Barbieri, G. G. Ross, and A. Strumia, hep-ph/9906470; 
N. Haba and N. Okamura, hep-ph/9906481.
\bibitem{8} See for example J. N. Bahcall, P. I. Krastev, and A. Yu. Smirnov, 
Phys. Rev. {\bf D58}, 096016 (1998).
\bibitem{9} L. Wolfenstein, Phys. Rev. {\bf D17}, 2369 (1978); S. P. Mikheyev 
and A. Yu. Smirnov, Sov. J. Nucl. Phys. {\bf 42}, 913 (1986).
\bibitem{10} L. Baudis {\it et al.}, Phys. Lett. {\bf B407}, 219 (1997); 
hep-ex/9902014.
\bibitem{11} E. Gawiser and J. Silk, Science {\bf 280}, 1405 (1998); J. R. 
Primack and M. A. K. Gross, astro-ph/9810204; K. S. Babu, R. K. Schaefer, and 
Q. Shafi, Phys. Rev. {\bf D53}, 606 (1996).
\end{thebibliography}

\end{document}